\documentclass[]{JHEP3}   % 10pt is ignored!
 \preprint{  }

\usepackage{epsfig,multicol}
\usepackage{amsmath}
\usepackage{graphics}
\usepackage{graphicx}
\usepackage{dcolumn}
\usepackage{amssymb}
\usepackage{appendix}

\title{Extension of non-minimal derivative coupling theory and Hawking radiation in black-hole
spacetime}
\author{Chikun Ding and Changqing Liu\thanks{Email: dingchikun@163.com, lcqliu2562@163.com}\\
 Department of Physics and Information Engineering, \\
Hunan Institute of Humanities Science and Technology,\\ Loudi, Hunan
417000, P. R. China}

\author{ Jiliang  Jing and Songbai Chen\thanks{Email:
jljing@hunnu.edu.cn, csb3752@163.com}\\
 Institute of Physics and  Department of Physics, \\
Hunan Normal University,\\ Changsha, Hunan 410081, P. R. China\\
Key Laboratory of Low Dimensional Quantum
Structures \\ and Quantum Control of Ministry of Education,\\ Hunan
Normal University, \\Changsha, Hunan 410081, People's Republic of
China}

\abstract{ We extend the non-minimal derivative coupling theory
model to dynamical gravity and use it to
 study the greybody factor and Hawking radiation in the background of the slowly rotating
Kerr-Newman black hole.
 Our
results show that both the absorption probability and luminosity of Hawking
radiation of the
scalar field increase with the coupling.
 Moreover, we also
find that for the weak coupling $\eta<\eta_c$, the absorption
probability and luminosity of Hawking radiation decrease when the
black hole's Hawking temperature decreases. On the other hand
 for stronger coupling $\eta>\eta_c$,
the absorption probability and luminosity of Hawking radiation
increase when the black hole's Hawking temperature decreases. This
feature is similar to the Hawking radiation in a $d$-dimensional
static spherically-symmetric black hole surrounded by quintessence
\cite{chensong}.
 }


\begin{document}

\section{Introduction}Scalar fields in General Relativity has been a topic of great interest in the latest years.
One of the main reasons is that the models with
scalar fields are relatively simple, which allows us to probe the
detailed features of the more complicated physical system. In
cosmology, scalar fields can be considered as candidate (inflaton, quintessence,
phantom fields, etc.) to explain
the inflation of the early Universe \cite{inf1} and the accelerated
expansion of the current Universe \cite{1a,2a,3a}. In the Standard
Model of particle physics, the scalar field presents as the Higgs
boson \cite{Hig}, which would help to explain the origin of mass in
the Universe. Moreover, it has been found that scalar field plays
the important roles in other fundamental physical theories, such as,
Jordan-Brans-Dicke theory \cite{bd}, Kaluza-Klein compactification
theory \cite{KK} and superstring theory \cite{dila}, and so on.

In another side, including nonlinear terms of the various curvature tensors (Riemann, Ricci, Weyl) and
nonminimally coupled terms in the effective action of gravity has become a very
common trend from quantum field theory side and cosmology. These theories cover the $f(R)$ modified gravity, the Gausss-Bonnet gravity, the tachyon, dilaton, and so on. The nonminimal coupling between scalar field and higher order
terms in the curvature (the so-called ``scalar-tensor" theory) naturally give rise to inflationary solutions improve the early inflationary models and could
contribute to solve the dark matter problem.
 The new coupling between the derivative of scalar field and the
spacetime curvature may appear firstly in some Kaluza-Klein theories
\cite{kkc,kkc1,kkc2}. Amendola \cite{AL1} considered the most
general theory of gravity with the Lagrangian linear in the Ricci
scalar, quadratic in $\psi$, in which the coupling terms have the
forms as follows
\begin{eqnarray}
R\partial_{\mu}\psi\partial^{\mu}\psi,\;\;
R_{\mu\nu}\partial^{\mu}\psi\partial^{\nu}\psi,\;\;
R\psi\nabla^2\psi, \;\;
R_{\mu\nu}\psi\partial^{\mu}\psi\partial^{\nu}\psi,\;\;\partial_{\mu}R\partial^{\mu}\psi,
\;\;\nabla^2R\psi.
\end{eqnarray}
And then he studied the dynamical evolution of the scalar field in
the cosmology by considering only the derivative coupling term
$R_{\mu\nu}\partial^{\mu}\psi\partial^{\nu}\psi$ and obtained some
analytical inflationary solutions \cite{AL1} . Capozziello
\textit{et al.} \cite{AL3} investigated a more general model of
containing coupling terms $R\partial_{\mu}\psi\partial^{\mu}\psi$
and $R_{\mu\nu}\partial^{\mu}\psi\partial^{\nu}\psi$, and found that
the de Sitter spacetime is an attractor solution in the model.
Recently, Daniel and Caldwell \cite{AL4}  obtained the constraints
on the theory with the derivative coupling term of
$R_{\mu\nu}\partial^{\mu}\psi\partial^{\nu}\psi$ by Solar system
tests. In general, a theory with derivative couplings could lead to
that both the Einstein equations and the equation of motion for the
scalar are the fourth-order differential equations. However, Sushkov
\cite{AL5} studied recently the model in which the kinetic term of
the scalar field only coupled with the Einstein's tensor and found
that the equation of motion for the scalar field can be reduced to
second-order differential equation. This means that the theory is a
``good" dynamical theory from the point of view of physics. Gao
\cite{g1} investigated the cosmic evolution of a scalar field with
the kinetic term coupling to more than one Einstein's tensors and
found the scalar field presents some very interesting characters. He
found that the scalar field behaves exactly as the pressureless
matter if the kinetic term is coupled to one Einstein's tensor and
acts nearly as a dynamic cosmological constant if it couples with
more than one Einstein's tensors. The similar investigations have
been considered in Refs.\cite{g2,g3}. These results will excite more
efforts to be focused on the study of the scalar field coupled with
tensors in the more general cases.

In this paper we will extend this non-minimal derivative coupling
theory model to dynamical gravity and use it to
 study the greybody factor and Hawking radiation in the background of the slowly rotating
Kerr-Newman black hole.

Since black hole is another fascinating object in modern physics, it
is of interest to extend the study the properties of the scalar
field when it is kinetically coupled to the Einstein's tensors in the
background of a black hole. This extension to Reissner-Nordstr\"{o}m black hole spacetime is studied by S. Chen {\it et al} \cite{songbai}. In this paper, we will investigate the
greybody factor and Hawking radiation of the scalar field coupling
 to the Einstein's tensor $G^{\mu\nu}$ in the slowly rotating
Kerr-Newman black hole spacetime. We find that the presence of the
coupling terms enhances both the absorption probability and
luminosity of Hawking radiation of the scalar field in the black
hole spacetime.  Moreover, we also find that for the weak coupling
$\eta<\eta_c$, the absorption probability and luminosity of Hawking
radiation decrease when the black hole's Hawking temperature
decreases; while for stronger coupling $\eta>\eta_c$, the absorption
probability and luminosity of Hawking radiation increase on the
contrary when the black hole's Hawking temperature decreases. This
feature is similar to the Hawking radiation in a $d$-dimensional
static spherically-symmetric black hole surrounded by quintessence
\cite{chensong}, i.e. when $0<-\omega_q<(d-3)/(d-1)$, Hawking
temperature decreases and the luminosity of Hawking radiation both
in the bulk and on the brane decreases naturally; when
$(d-3)/(d-1)<-\omega_q<1$, Hawking temperature still deceases, but
the luminosity of Hawking radiation both in the bulk and on the
brane increases conversely.

The paper is organized as follows: in the following section we will
introduce the action of a scalar field coupling to Einstein's tensor
and derive its master equation in the slowly rotating Kerr-Newman
black hole spacetime. In Sec. 3, we obtain the expression of the
absorption probability in the low-energy limit by using the matching
technique. In section 4, we will calculate the absorption
probability and the luminosity of Hawking radiation for the coupled
scalar field. In section 5, we will include and discuss our
conclusions. Appendix is devoted to the extension of the non-minimal
derivative coupling theory model to the dynamical gravity.

\section{Master equation with non-minimal derivative coupling in
the slowly rotating black hole spacetime}

Let us consider the action of the scalar field coupling to the
Einstein's tensor $G^{\mu\nu}$ in the curved spacetime \cite{AL5},
\begin{eqnarray}\label{action}
S=\int d^4x \sqrt{-g}\bigg[\frac{R}{16\pi
G}+\frac{1}{2}\partial_{\mu}\psi\partial^{\mu}\psi+\frac{\eta}{2}
G^{\mu\nu}\partial_{\mu}\psi\partial_{\nu}\psi\bigg].
\end{eqnarray}
The coupling between Einstein's tensor $G^{\mu\nu}$ and the scalar
field $\psi$ is represented by
$\frac{\eta}{2}G^{\mu\nu}\partial_{\mu}\psi\partial_{\nu}\psi$,
where $\eta$ is the coupling constant with dimensions of
length-squared. In general, the presence of such a coupling term
brings some effects to the original metric of the background.
However, we can treat the scalar filed as a perturbation so that the
backreaction effects on the background can be ignored, and then we
can study the effects of the coupling constant $\eta$ on the
greybody factor and Hawking radiation of the scalar filed in a black
hole spacetime. In addition, to avoid the kinetic instability for
which $g^{00}+\eta G^{00}>0\;(g^{00}<0$ is assumed) , we set the
coupling constant to be positive, $\eta>0$.

Varying the action with respect to $\psi$, one can obtain the
modified Klein-Gordon equation
\begin{eqnarray}
\frac{1}{\sqrt{-g}}\partial_{\mu}\bigg[\sqrt{-g}\bigg(g^{\mu\nu}+\eta
G^{\mu\nu}\bigg)\partial_{\nu}\psi\bigg] =0,\label{WE}
\end{eqnarray}
which is a second order differential equation. Obviously, all the
components of the tensor $G^{\mu\nu}$ vanish in the Kerr
black hole spacetime because it is the vacuum solution of the
Einstein's field equation. Thus, we cannot probe the effect of the
coupling term on the greybody factor and Hawking radiation in the
Kerr black-hole background. The simplest rotating black hole
with the non-zero components of the tensor $G^{\mu\nu}$ is
Kerr-Newman one. In this paper, we consider a slowly rotating Kerr-Newman
 black hole, whose element reads \cite{shir}
\begin{eqnarray}
&~& ds^2 = -\frac{\Delta}{r^2} dt^2 -
\frac{2(2Mr-Q^2)a \sin^2\theta}{r^2}\,dt \, d\varphi
+\frac{r^2}{\Delta}\,dr^2
 +r^2d\Omega^2+\mathcal{O}(a^2)
\label{m1}
\end{eqnarray}
with
\begin{equation}
\Delta = r^2  -2Mr+Q^2\,,
\label{Delta}
\end{equation}
where $M$, $a$, $Q$ are the mass, angular momentum and charge of the black hole. The
Einstein's tensor $G^{\mu\nu}$ for the metric (\ref{m1}) has a form
\begin{eqnarray}
G^{\mu\nu}= \frac{Q^2}{r^4} \left(\begin{array}{cccc}
 -\frac{r^2}{\Delta}&0&0&
 -\frac{a}{r^2\Delta}(r^2+\Delta)\\
 0&\frac{\Delta}{r^2}&0&0\\
 0&0&-\frac{1}{r^2}&0\\
 -\frac{a}{r^2\Delta}(r^2+\Delta)&0&0&
 -\frac{1}{r^2\sin^2\theta}
\end{array}\right).
\end{eqnarray}
Adopting to the spherical harmonics
%%%%%%%%
\begin{equation}
\phi(t,r,\theta,\varphi)= e^{-i\omega t}\,e^{i m \varphi}\,R_{\omega \ell m}(r)
\,T^{m}_{\ell}(\theta, a \omega)\,,
\end{equation}
%%%%%%%%%
 we can
 obtain the radial part of the equation (\ref{WE})
%%%%%%%%%
\begin{equation}
\frac{d}{dr}\biggl[\Delta\big(1+\frac{\eta Q^2}{r^4}\big)\,\frac{d R_{\omega \ell m}}{dr}\biggr]+
\left[\frac{r^2\omega(r^2\omega-2am)}{\Delta}\big(1+\frac{\eta Q^2}{r^4}\big) -[l(l+1)-2am\omega]\big(1-\frac{\eta Q^2}{r^4}\big)
\right]R_{\omega \ell m}=0\,. \label{radial}
\end{equation}
%%%%%%%%%%
Clearly, the radial equation (\ref{radial}) contains the
coupling constant $\eta$, which means that the presence of the
coupling term will change the evolution of the scalar field in the
Kerr-Newman black hole spacetime.

The solution of the radial function $R_{\omega \ell m}(r)$ will help us to obtain
the absorption probability $|A_{\ell m}|^2$ and the luminosity of Hawking
radiation for a scalar field coupling with Einstein's tensor in the slowly rotating
Kerr-Newman black hole spacetime.

\section{Greybody Factor in the Low-Energy Regime}

In order to study the effects of the coupling constant $\eta$ on the
absorption probability $|A_{\ell m}|^2$ and the luminosity of
Hawking radiation of a scalar field in the background spacetime, we
must first get an analytic solution of the radial equation
(\ref{radial}). In general, it is very difficult. However, as in
ref.\cite{Kanti,Kan1,Kan2,Kan3,Kan4,Kan5,Kan6,Haw3,Haw4,Haw5}, we
can provide an approximated solution of the radial equation
(\ref{radial}) by employing the matching technique. Firstly, we must
derive the analytic solutions in the near horizon ($r\simeq r_+$)
and far-field $(r\gg r_+)$ regimes in the low-energy limit. Finally,
we smoothly match these two solutions in an intermediate region. In
this way, we can construct a smooth analytical solution of the
radial equation valid throughout the entire spacetime.

Now, we focus on the near-horizon regime and perform the following
transformation of the radial variable as in Refs.
\cite{Haw3,Haw4,Haw5}
%%%%%%%%%%
\begin{eqnarray}
r \rightarrow f(r) = \frac{\Delta(r)}{r^2} \,\,\Longrightarrow\,
\frac{d f}{dr}=(1-f)\frac{\mathcal{A}(r)}{r}\,,
\end{eqnarray}
%%%%%%%%%%
with
\begin{eqnarray}
\mathcal{A}=1-\frac{Q^2}{2Mr-Q^2}.\label{ax}
\end{eqnarray}
The equation (\ref{radial}) near the horizon $(r\sim r_+)$ can be
rewritten as
\begin{eqnarray}
f(1-f)\frac{d^2R(f)}{d f^2}+(1-D_*f)\frac{d R(f)}{d f}
+\bigg[\frac{K^2_*}{A_*^2(1-f)f}
-\frac{\Lambda^m_\ell}{A_*^2(1-f)}\bigg(\frac{1-\eta_*
Q_*^2}{1+\eta_*
Q_*^2}\bigg)\bigg]R(f)=0,\label{r1}
\end{eqnarray}
where \footnote{In order to solve the mathematical equation, we
should let all coefficients in it dimensionless, so that we define
these quantities.  Usually we let the numerical value of $r_+$ equal
to unity. However in this paper, we find that to let the numerical
value of $2M$ equal to 1 is more convenient.}
\begin{eqnarray}
&&a_*=a/r_+,\; Q_*=Q/r_+,\; \eta_*=\eta/r_+^2,\; K_*=\omega  r_+-a_*m,\;A_*=1-Q_*^2,\nonumber\\&&
\Lambda^m_\ell=l(l+1)-2ma\omega,\;D_*=1-\frac{2Q_*^2}{(1-Q_*^2)^2}+\frac{4\eta_* Q_*^2}
{(1-Q_*^2)[1+\eta_* Q_*^2]}.\label{kx}
\end{eqnarray}
Making the field redefinition $R(f)=f^{\alpha}(1-f)^{\beta}F(f)$,
one can find that the equation (\ref{r1}) can be rewritten as a form
of the hypergeometric equation
\begin{eqnarray}
f(1-f)\frac{d^2F(f)}{d f^2}+[c-(1+\tilde{a}+b)f]\frac{d F(f)}{d
f}-\tilde{a}b F(f)=0,\label{near2}
\end{eqnarray}
with
\begin{eqnarray}
\tilde{a}=\alpha+\beta+D_*-1,\;\;\;\;\;\;\;\;\;\;
b=\alpha+\beta,\;\;\;\;\;\;\;\;\;\;\;\;\; c=1+2\alpha.
\end{eqnarray}
Considering the constraint coming from coefficient of $F(f)$, one
can easy to obtain that the power coefficients $\alpha$ and $\beta$
satisfy
\begin{eqnarray}
\alpha^2+\frac{K^2_*}{A_*^2}=0,
\end{eqnarray}
and
\begin{eqnarray}
\beta^2+\beta(D_*-2)+\frac{1}{A_*^2}\bigg[K^2_*
-\Lambda^m_\ell\bigg(\frac{1-\eta_*
Q_*^2}{1+\eta_*
Q_*^2}\bigg)\bigg]=0,
\end{eqnarray}
respectively. These two equations admit that the parameters $\alpha$
and $\beta$ have the forms
\begin{eqnarray}
&&\alpha_{\pm}=\pm \frac{iK_*}{A_*},\\
&&\beta_{\pm}=\frac{1}{2}\bigg\{(2-D_*)\pm\sqrt{(D_*-2)^2-\frac{4}{A_*^2}\bigg[K^2_*
-\Lambda^m_\ell\bigg(\frac{1-\eta_* Q_*^2}{1+\eta_*
Q_*^2}\bigg)\bigg]} \;\bigg\}.\label{bet}
\end{eqnarray}
The general solution of Eq. (\ref{near2}) is
 \begin{eqnarray}
R_{NH}(f)=A_-f^{\alpha}(1-f)^{\beta}F(\tilde{a}, b, c; f)+A_+
f^{-\alpha}(1-f)^{\beta}F(\tilde{a}-c+1, b-c+1, 2-c; f),
\end{eqnarray}
where $A_+, A_-$ are arbitrary constants. Near the horizon,
$r\rightarrow r_h$ and $f\rightarrow0$, the solution has the form
 \begin{eqnarray}
R_{NH}(f)=A_-f^{\alpha_\mp}+A_+f^{\alpha_\pm}.
\end{eqnarray}
Imposing the boundary condition that no outgoing mode exists near
the horizon, we are forced to set either $A_-=0$ or $A_+=0$,
depending on the choice for $\alpha_\pm$. Here we choose
$\alpha=\alpha_-$ and $A_+=0$. The sign of $\beta$ will be decided
by the criterion for the convergence of the hypergeometric function
$F(\tilde{a}, b, c; f)$, i.e. Re$(c-\tilde{a}-b)>0$, which demands
that we choose $\beta=\beta_-$ \cite{Haw3,Haw4,Haw5}.
 Thus the asymptotic solution near horizon has the
form
\begin{eqnarray}
R_{NH}(f)=A_-f^{\alpha}(1-f)^{\beta}F(\tilde{a}, b, c; f).
\end{eqnarray}

Let us now to stretch smoothly the near horizon solution to the
intermediate zone. We can make
use of the property of the hypergeometric function \cite{mb} and
change its argument in the near horizon solution from $f$ to $1-f$
\begin{eqnarray}
R_{NH}(f)&=&A_-f^{\alpha}(1-f)^{\beta}\bigg[\frac{\Gamma(c)\Gamma(c-\tilde{a}-b)}
{\Gamma(c-\tilde{a})\Gamma(c-b)}
F(\tilde{a}, b, \tilde{a}+b-c+1; 1-f)\nonumber\\
&+&(1-f)^{c-\tilde{a}-b}\frac{\Gamma(c)\Gamma(\tilde{a}+b-c)}{\Gamma(\tilde{a})\Gamma(b)}
F(c-\tilde{a}, c-b, c-\tilde{a}-b+1; 1-f)\bigg].\label{r2}
\end{eqnarray}
As $r\gg r_+$, the function $(1-f)$ can be approximated as
\begin{eqnarray}
1-f=\frac{2Mr-Q^2}{r^2}\simeq \frac{2M}{r},
\end{eqnarray}
and then the near horizon solution (\ref{r2}) can be simplified
further to
\begin{eqnarray}
R_{NH}(r)\simeq C_1r^{-\beta}+C_2r^{\beta+D_*-2}\label{rn2},
\end{eqnarray}
with
\begin{eqnarray}
C_1=A_-(2M)^{\beta}
\frac{\Gamma(c)\Gamma(c-\tilde{a}-b)}{\Gamma(c-\tilde{a})\Gamma(c-b)},\label{rn3}
\end{eqnarray}
\begin{eqnarray}
C_2=A_-(2M)^{-(\beta+D_*-2)}\frac{\Gamma(c)\Gamma(\tilde{a}+b-c)}{\Gamma(\tilde{a})\Gamma(b)}.\label{rn4}
\end{eqnarray}

Nextly in order to obtain a solution in the far field region, we expand the
wave equation (\ref{radial}) as a power series in $1/r$ and keep
only the leading terms
\begin{eqnarray}
\frac{d^2R_{FF}(r)}{dr^2}+\frac{2}{r}\frac{dR_{FF}(r)}{d
r}+\bigg(\omega^2-\frac{l(l+1)}{r^2}\bigg)R_{FF}(r)=0.
\end{eqnarray}
This is the usual Bessel equation. Thus the solution of the radial
master equation (\ref{radial}) in the far-field limit can be
expressed as
\begin{eqnarray}
R_{FF}(r)=\frac{1}{\sqrt{r}}\bigg[B_1J_{\nu}(\omega\;r)+B_2Y_{\nu}
(\omega\;r)\bigg],\label{rf}
\end{eqnarray}
where $J_{\nu}(\omega\;r)$ and $Y_{\nu}(\omega\;r)$ are the first
and second kind Bessel functions, $\nu=l+1/2$. $B_1$ and
$B_2$ are integration constants. In order to stretch the far-field
solution (\ref{rf}) towards small radial coordinate, we take the
limit $r\rightarrow 0$ and obtain
\begin{eqnarray}
R_{FF}(r)\simeq\frac{B_1(\frac{\omega\;r}{2})^{\nu}}{\sqrt{r}\;\Gamma(\nu+1)}
-\frac{B_2\Gamma(\nu)}{\pi
\sqrt{r}\;(\frac{\omega\;r}{2})^{\nu}}.\label{rfn2}
\end{eqnarray}
In the low-energy and low-angular momentum limit $(\omega r_+)^2\ll1$ and
$(a/r_+)^2\ll1$, the two power coefficients in
Eq.(\ref{rn2}) can be approximated as
\begin{eqnarray}
-\beta &\simeq &l + {\cal O}(\omega^2,a^2,a\omega), \\
(\beta+D_*-2)&\simeq &-(l+1)+ {\cal O}(\omega^2,a^2,a\omega).
\end{eqnarray}
 By using the above results, one can easily show that both Eqs. (\ref{rn2}) and (\ref{rfn2}) reduce to power-law expressions with the same power
coefficients, $r^l$ and $r^{-(l+1)}$. By matching the corresponding coefficients between
Eqs. (\ref{rn2}) and (\ref{rfn2}), we can obtain two relations
between $C_1,\;C_2$ and $B_1,\;B_2$. Removing $A_-$, we can obtain
the ratio between the coefficients $B_1,\; B_2$
\begin{eqnarray}
B\equiv\frac{B_1}{B_2}&=&-\frac{1}{\pi}\bigg[\frac{1}{\omega M}\bigg]^{2l+1}
\nu\Gamma^2(\nu)
\nonumber\\
&\times&\;\frac{
\Gamma(c-\tilde{a}-b)\Gamma(\tilde{a})\Gamma(b)}{\Gamma(\tilde{a}+b-c)\Gamma(c-\tilde{a})\Gamma(c-b)}.
\label{BB}
\end{eqnarray}
In the asymptotic region $r\rightarrow \infty$, the solution in the
far-field can be expressed as
\begin{eqnarray}
R_{FF}(r)&\simeq &
\frac{B_1+iB_2}{\sqrt{2\pi\;\omega}\;r}e^{-i\omega\;r}+
\frac{B_1-iB_2}{\sqrt{2\pi\;\omega}\;r}e^{i\omega\;r}\\
&=& A^{(\infty)}_{in}\frac{e^{-i\omega\;r}}{r}
+A^{(\infty)}_{out}\frac{e^{i\omega\;r}}{r}.\label{rf6}
\end{eqnarray}
The absorption probability can be calculated by
\begin{eqnarray}
|\mathcal{A}_{\ell m}|^2=1-\bigg|\frac{A^{(\infty)}_{out}}{A^{(\infty)}_{in}}\bigg|^2
=1-\bigg|\frac{B-i}{B+i}\bigg|^2=\frac{2i(B^*-B)}{BB^*+i
(B^*-B)+1}.\label{GFA}
\end{eqnarray}
Inserting the expression of $B$ (\ref{BB}) into Eq.(\ref{GFA}), we
can probe the properties of absorption probability for the scalar
field coupled with Einstein's tensor in the slowly rotating black hole
spacetime in the low-energy limit.

\section{The absorption probability and Hawking radiation with
non-minimal derivative coupling}

We are now in a position to calculate the absorption probability and
discuss Hawking radiation of a scalar field coupling to Einstein's
tensor in the background of a slowly rotating Kerr-Newman black hole.

In Fig. 1, we fix the coupling constant $\eta$, and angular momentum
$a$, and plot the change of the absorption probability of a scalar
particle with the charge $Q$ for the first partial waves ($\ell=0$)
in the slowly rotating Kerr-Newman black hole. One can easily see
that for the smaller $\eta$ the absorption probability $A_{\ell=0}$
decreases with the charge $Q$ of the black hole, which is similar to
that for the usual scalar field without coupling to Einstein's
tensor. However, for the larger $\eta$, the absorption probability
$A_{\ell=0}$ increases as the charge $Q$ increases. These properties
mean that the stronger coupling between the scalar field and
Einstein's tensor changes the properties of the absorption
probability of scalar field in the black hole spacetime. In Fig.2,
we also find that the absorption probability increases with the
increase of the coupling constant $\eta$ for fixed values of charge
$q=0.3$, and $a=0.1$.
\begin{figure}[ht]\label{fig1}
\begin{center}
\includegraphics[width=8.0cm]{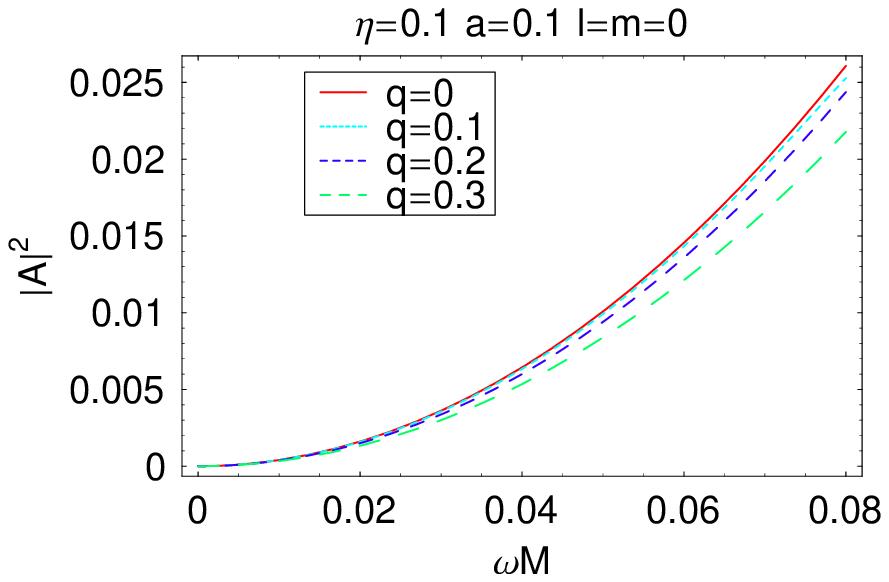}\;\;\;\;\includegraphics[width=8.0cm]{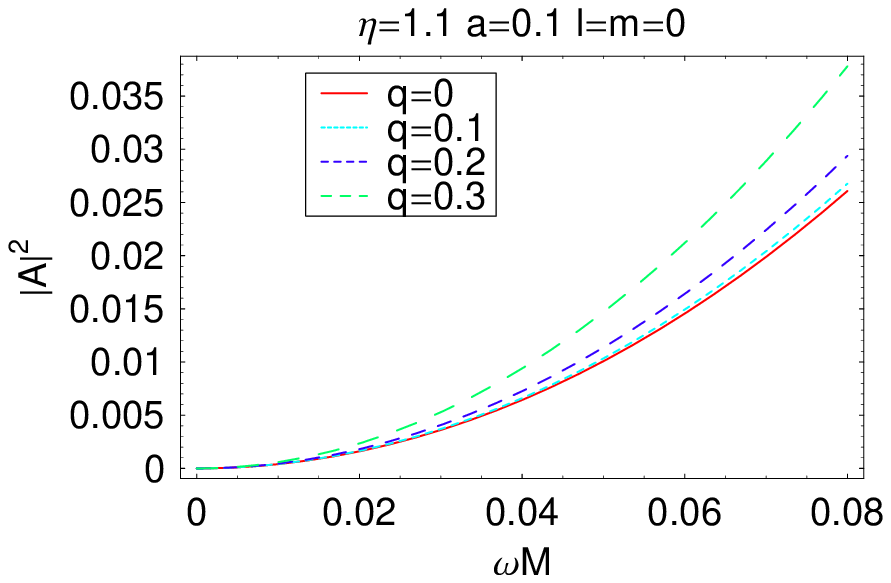}
\caption{Variety of the absorption probability $|A_{\ell m}|^2$ of a
 scalar field with the charge $Q$ and angular momentum in the slowly rotating Kerr-Newman black hole for
 fixed $\ell=m=0$.  The coupling constant $\eta$ is set by $\eta=0.1$ in the left and by
   $\eta=1.1$ in the right.
}
\end{center}
\end{figure}
\begin{figure}[ht]\label{fig2}
\begin{center}
\includegraphics[width=8.0cm]{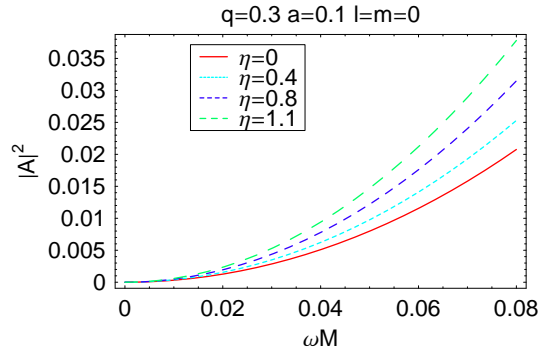}
\caption{The dependence of the absorption probability $|A_{\ell
m}|^2$ of a scalar field on the coupling constant $\eta$ in the
slowly rotating Kerr-Newman black hole for fixed $\ell=m=0$ and
$Q=0.3$. }
 \end{center}
\end{figure}
In Fig. 3, we show the Hawking temperature of the slowly rotating
Kerr-Newman black hole with various charge $Q$. Therefore,
for weak coupling, when Hawking temperature decreases, the greybody
 factor decreases naturally, but for stronger coupling the greybody factor on the contrary increases.
\begin{figure}[ht]\label{fig11}
\begin{center}
\includegraphics[width=8.0cm]{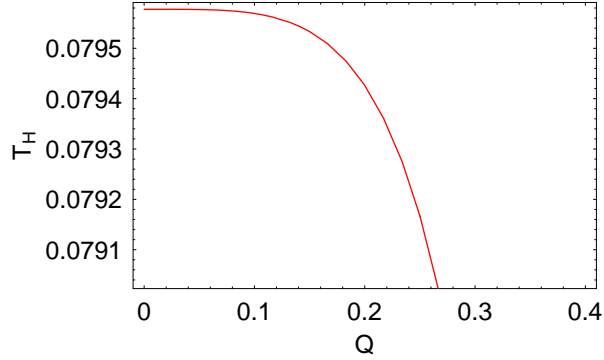}
\caption{The Hawking temperature of the slowly rotating Kerr-Newman
black hole with various charge $Q$. }
 \end{center}
\end{figure}

The above results about the absorption
probability also hold true for other values of $\ell, m$. In this case, there
has superradiation region when $m=1, 2, \cdots, \ell$, which is similar to \cite{Haw3}. It is shown in
Fig. 4, in which we plotted the dependence of the absorption
probability on the angular index $\ell$ and $m$ with different $\eta$, $a$ and $Q$. From the above two figures in Fig. 4, we can obtain that the influence of charge $Q$ on the usual radiation ($m=-1, 0$) is similar to that on the first partial wave. While for the super-radiation, the charge enhance it
 both for weak and strong coupling. And the angular momentum $a$ enhance the usual radiation ($m=-1$) and the super-radiation ($m=1$) both for weak and strong coupling.
Moreover, we see the suppression
of $|A_{\ell m}|^2$ as the values of the angular index increase. This means
that the first partial wave dominates over all others in the
absorption probability. It is similar to that of the scalar field
without coupling to Einstein's tensor as shown in
refs.\cite{Kanti,Kan1,Kan2,Kan3,Kan4,Kan5,Kan6,Haw3,Haw4,Haw5}.
\begin{figure}[ht]\label{fig3}
\begin{center}
\includegraphics[width=8.0cm]{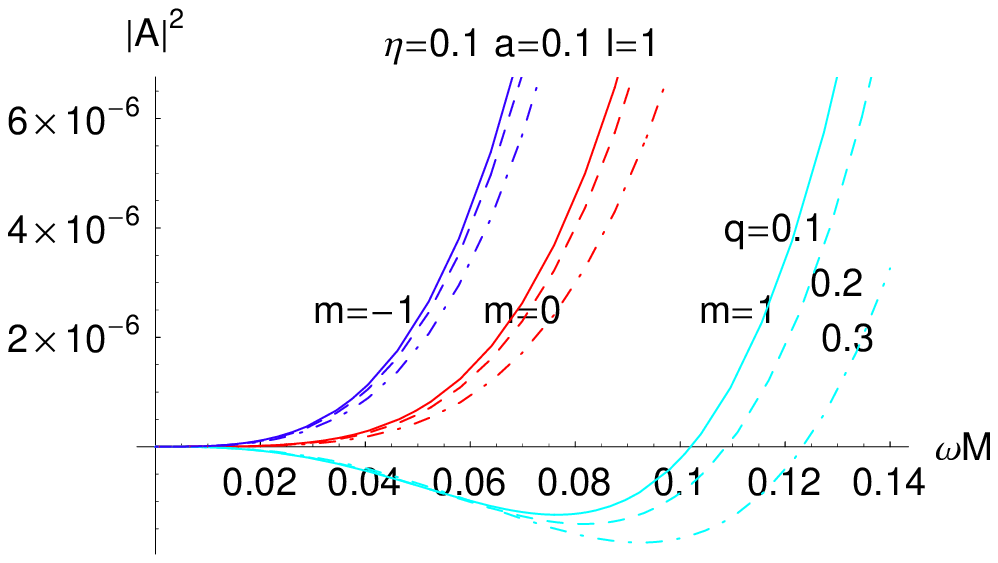}\includegraphics[width=8.0cm]{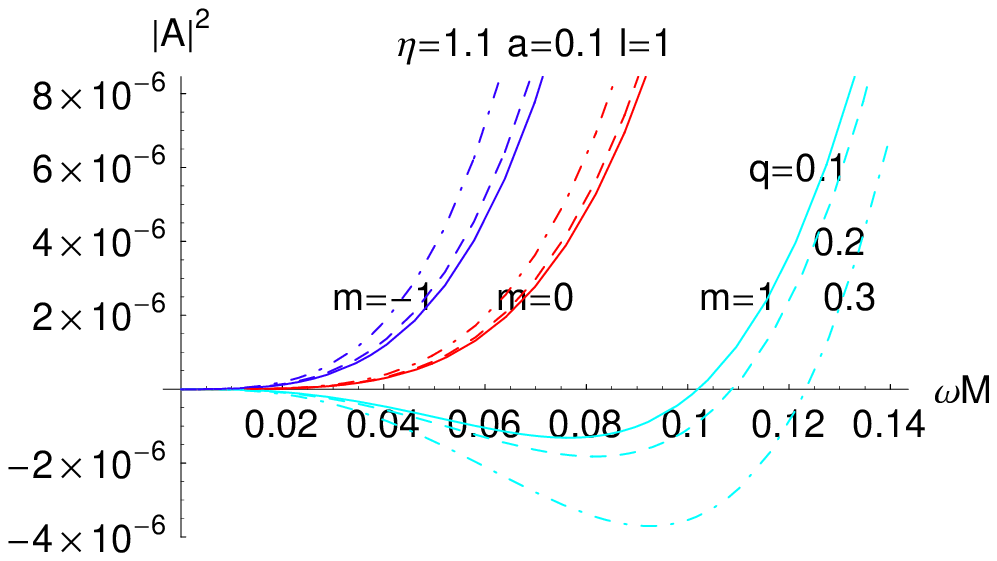}
\includegraphics[width=8.0cm]{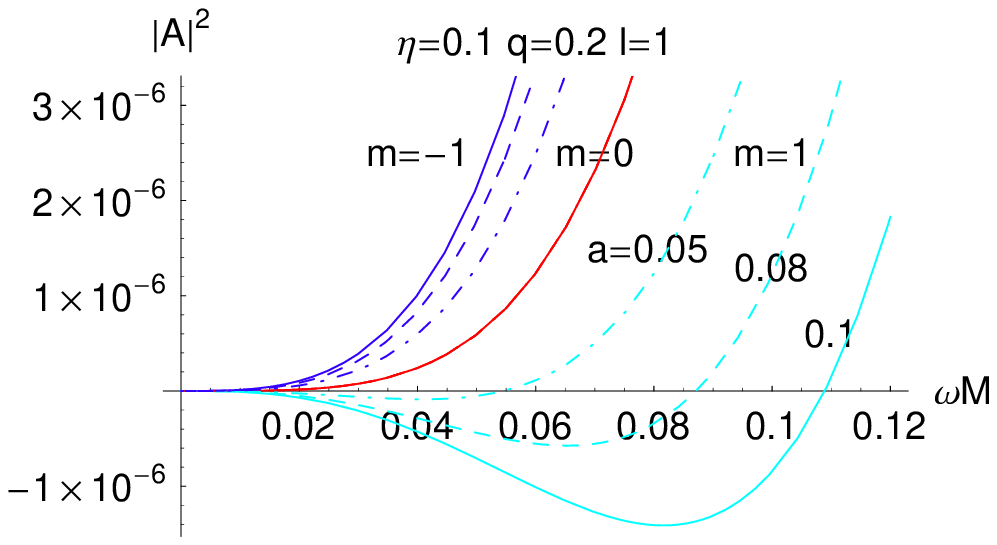}
\includegraphics[width=8.0cm]{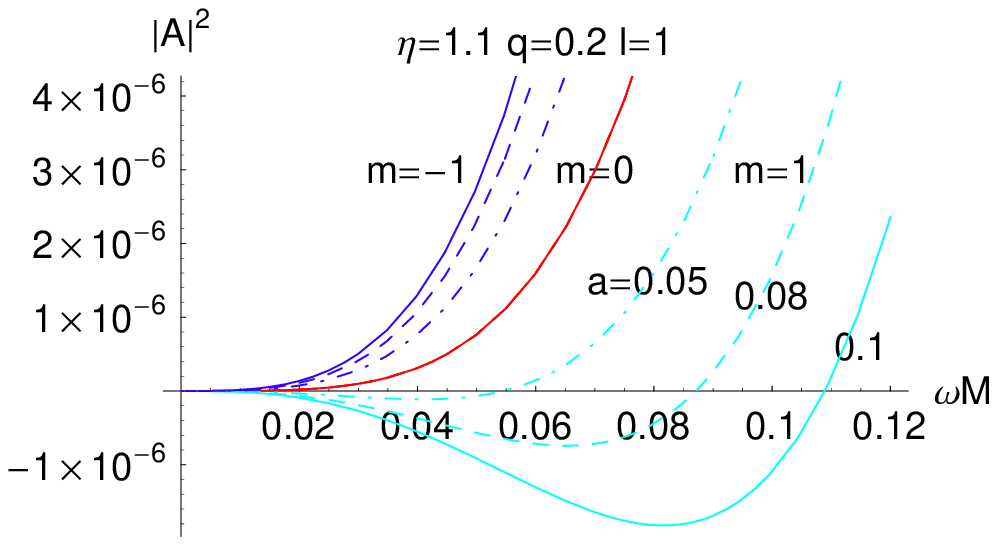}
\caption{Variety of the absorption probability $|A_{\ell m}|^2$ of a
 scalar field with the charge $Q$ and angular momentum $a$ in the slowly rotating Kerr-Newman black hole spacetime for
 fixed $\ell=1, m=1, 0, -1$. The coupling constant $\eta$ is set by $\eta=0.1$ on the left and $\eta=1.1$ on the right. In the above two figures, the solid lines represent $Q=0.1$, the dashed lines represent $Q=0.2$, the dashed-dotted lines represent $Q=0.3$; in the lower two figures, the solid lines represent $a=0.1$, the dashed lines represent $a=0.08$, the dashed-dotted lines represent $a=0.05$.
  }
\end{center}
\end{figure}

Now let us turn to study the luminosity of the Hawking radiation for
the mode $\ell=m=0$ which plays a dominant role in the greybody factor.
Performing an analysis similar to that in \cite{Haw3,Haw4,Haw5}, we
can obtain that the greybody factor (\ref{GFA}) in the low-energy
limit has a form
\begin{eqnarray}
|A_{\ell=0}|^2&\simeq& \frac{4\omega^2
r^2_+(r^2_++Q^2)}{(r^2_+-Q^2)(2-D_{\ast})}.\label{GFA1}
\end{eqnarray}
Combining it with Hawking temperature $T_H$ \cite{ding} of the
Kerr-Newman  black hole,
\begin{eqnarray}
T_H=\frac{r_+-M}{2\pi(r_+^2+a^2)}=\frac{r_+^2-a^2-Q^2}{4\pi r_+(r_+^2+a^2)}=\frac{r_+^2-Q^2}{4\pi r_+^3}+\mathcal{O}(a^2),
\end{eqnarray}
 the luminosity of the Hawking
radiation for the scalar field with coupling to Einstein's tensor is
given by
\begin{eqnarray}
L&=&\int^{\infty}_0\frac{d\omega}{2\pi}
|A_{\ell=0}|^2\frac{\omega}{e^{\;\omega/T_{H}}-1} \label{LHK}.
\end{eqnarray}
The integral expressions above are just for the sake of completeness
by writing the integral range from $0$ to infinity. However, as our
analysis has focused only in the low-energy regime of the spectrum,
an upper cutoff will be imposed on the energy parameter so that the
low-energy conditions $\omega\ll T_H$ and $\omega r_+\ll 1$ are
satisfied. In the low-energy limit, the luminosity of the Hawking
radiation for the mode $\ell=0$ can be approximated as
\begin{eqnarray}
L\approx\frac{2\pi^3}{15}GT^4_H,
\end{eqnarray}
with
\begin{eqnarray}
G=\frac{
r^2_+(r^2_++Q^2)}{(r^2_+-Q^2)(2-D_{\ast})}. \label{gs}
\end{eqnarray}
In Fig. 5 and 6, we show the dependence of the luminosity of Hawking
radiation on the charge $Q$ and the coupling constant $\eta$,
respectively. From Fig. 5, one can easily obtain that with increase
of $Q$ the luminosity of Hawking radiation $L$ decreases for the
smaller $\eta$ and increases for the larger $\eta$. In other words, for the weak coupling, the luminosity of Hawking radiation decreases naturally when the black hole's Hawking temperature decreases; while for stronger coupling,
the luminosity of Hawking radiation increases on the contrary when the black hole's Hawking temperature decreases. This is similar
to the behavior of the absorption probability discussed previously. This feature is similar to the Hawking radiation in a $d$-dimensional static spherically-symmetric black hole surrounded by quintessence \cite{chensong}, i.e.
when $0<-\omega_q<(d-3)/(d-1)$, Hawking temperature deceases, the luminosity of Hawking radiation both in the bulk and
on the brane decrease naturally; when $(d-3)/(d-1)<-\omega_q<1$, Hawking temperature still deceases, the luminosity of Hawking radiation both in the bulk and
on the brane increase conversely.
In Fig. 6, we show that the luminosity of Hawking radiation $L$
increases monotonously with the coupling constant $\eta$ for the all
$Q$.
\begin{figure}[ht]\label{fig5}
\begin{center}
\includegraphics[width=8.0cm]{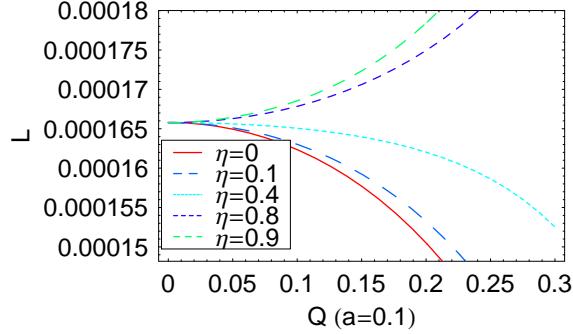}
\caption{Variety of the luminosity of Hawking radiation $L$ of
scalar particles with  the charge $Q$ and angular momentum $a$ in
the slowly rotating Kerr-Newman black hole for fixed $\ell=0$ and
different values of $\eta$.  }
\end{center}
\end{figure}
\begin{figure}[ht]\label{fig6}
\begin{center}
\includegraphics[width=8.0cm]{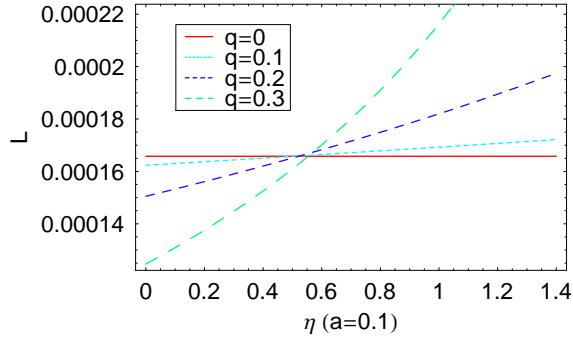}
\caption{The dependence of the luminosity of Hawking radiation $L$
of a scalar field on the coupling constant $\eta$ in the slowly
rotating Kerr-Newman black hole for fixed $\ell=0$ and different
values of $Q$  and angular momentum $a$. }
\end{center}
\end{figure}

From the Fig. 5, there should exist a critical coupling constant $\eta_c$, so in the table I, we list the critical coupling constant $\eta_c$ for different $Q$ by using equation $\partial L/\partial Q=0$. The results show that different charge $Q$ of the slowly rotating Kerr-Newman black hole correspond to different critical coupling constant $\eta_c$.
\begin{table}[!h]
\begin{center}
\begin{tabular}{cccc}
 \hline\hline\;\;\;\; $Q$ \;\;\;\; & \;\;\;\; $0.1$\;\;\;\;
& \;\;\;\;  $0.2$\;\;\;\;
 & \;\;\;\; $0.3$ \;\;\;\;
 \\
\hline
$\eta_c$& \;\;\;\;\;$0.510$\;\;\;\;\;  &
\;\;\;\; $0.541$\;\;\;\;\;
 & \;\;\;\;\;$0.600$\;\;\;\;\;
 \\
\hline
\end{tabular}\label{tab0}
\caption{The critical value of coupling constant $\eta_c$ for
different $Q$. We set $\ell=m=0$. }
\end{center}
\end{table}

\section{Summary and discussion}
In this paper, we have extend the non-minimal derivative coupling
theory model to dynamical gravity and
  studied the greybody factor and Hawking
radiation with a non-minimal derivative coupling between a scalar
field and the curvature in the background of the slowly rotating
Kerr-Newman black hole spacetime. We have found that the presence of
the coupling enhances  both the absorption probability and the
luminosity of Hawking radiation of the scalar field in the black
hole spacetime. Moreover, we also find that for the weak coupling
$\eta<\eta_c$, the absorption probability and the luminosity of
Hawking radiation decrease when the black hole's Hawking temperature
decreases; while for stronger coupling $\eta>\eta_c$, the absorption
probability and the luminosity of Hawking radiation increase on the
contrary when the black hole's Hawking temperature decreases. This
feature is similar to the Hawking radiation in a $d$-dimensional
static spherically-symmetric black hole surrounded by quintessence
\cite{chensong}, i.e. when $0<-\omega_q<(d-3)/(d-1)$, Hawking
temperature decreases, the luminosity of Hawking radiation both in
the bulk and on the brane decrease naturally; when
$(d-3)/(d-1)<-\omega_q<1$, Hawking temperature still deceases, the
luminosity of Hawking radiation both in the bulk and on the brane
increase conversely.

 {\it Discussion:} This very amazing
and interesting similarity shows that the scalar field which
 is coupled to the Einstein's tensor of a black hole spacetime may behave as
  a ``dark energy model" in the cosmology. In the Ref. \cite{g1}, the authors point
  out from cosmological side
  that this scalar field which is coupled to the Einstein's tensor behaves exactly
  as pressureless matter (without a scalar potential), plays the role of both cold
  dark matter and dark energy (with a scalar potential). Whether
  there exists this scalar-tensor coupling or not is still an open question.
Because the action (\ref{action}) contains a dimension six operator,
which when considered at the effective theory level is an irrelevant
operator, and hence should not have any relevance on sub-Planckian
scales. All these issues need further studying.

\begin{acknowledgments}
This work was partially supported by the Scientific Research
Foundation for the introduced talents of Hunan Institute of
Humanities Science and Technology. J. Jing's work was partially
supported by the National Natural Science Foundation of China under
Grant No.10675045, No.10875040 and No.10935013; 973 Program Grant
No. 2010CB833004 and the Hunan Provincial Natural Science Foundation
of China under Grant No.08JJ3010. S. Chen's work was  partially
supported by the National Natural Science Foundation of China under
Grant No.10875041,  the Program for Changjiang Scholars and
Innovative Research Team in University (PCSIRT, No. IRT0964) and the
construct program of key disciplines in Hunan Province.
\end{acknowledgments}

\begin{center}
{\bf APPENDIX}
\end{center}
\appendix\section{Extending the non-minimal derivative coupling theory model to dynamical
gravity}

In this appendix, we extend the non-minimal derivative coupling
model to the dynamical gravity. The metric tensor is
$g_{\mu\nu}=\eta_{\mu\nu}+h_{\mu\nu}$, where the components
$h_{\mu\nu}$ be much smaller than $1$ and $\eta_{\mu\nu}$ is the
metric tensor of Minkowski. Keeping the leading non-vanishing
contributions to the action, and dropping some boundary terms, we
get the following action,
\begin{eqnarray}
S&=&-\frac{1}{16\pi G}\int d^4x
\bigg[\frac{1}{2}h^{\mu\nu}D_{\mu\nu\rho\sigma}h^{\rho\sigma}\nonumber\\&&
+\int
d^4x\bigg[-\frac{1}{2}\Big(\eta^{\mu\nu}+\frac{1}{2}\eta^{\mu\nu}h-h^{\mu\nu}\Big)\bigg]
(\partial_{\mu}\psi)(\partial_{\nu}\psi)-\frac{\tilde{\eta}}{4}
(D^{\mu\nu}_{\rho\sigma}h^{\rho\sigma})(\partial_{\mu}\psi)(\partial_{\nu}\psi)\bigg],
\end{eqnarray}
 where here indices are raised/lowered with $
 \eta_{\mu\nu}$ and where
$D_{\mu\nu\rho\sigma}$ represents the well known graviton dynamical
operator,
\begin{eqnarray}
 D_{\mu\nu\rho\sigma}=\frac{1}{2}\bigg[
\partial_{(\rho}\eta_{\sigma)(\mu}\partial_{\nu)}-\frac{1}{2}\eta_{\mu(\rho}\eta_{\sigma)\nu}
\partial^2-\frac{1}{2}\eta_{\rho\sigma}\partial_{\mu}\partial_{\nu}
-\frac{1}{2}\eta_{\mu\nu}\partial_\rho\partial_\sigma+\frac{1}{2}\eta_{\mu\nu}
\eta_{\rho\sigma}\partial^2\bigg],
\end{eqnarray}
and $\tilde{\eta}$ is a coupling constant. Now varying the action
with respect to $h^{\mu\nu}$ and $\psi$ gives the following
equations of motion,
\begin{eqnarray}
 -\frac{1}{16\pi
G}D_{\mu\nu\rho\sigma}h^{\rho\sigma}+\frac{1}{2}(\partial_{\mu}\psi)(\partial_{\nu}\psi)
-\frac{1}{4}\eta_{\mu\nu}\eta^{\alpha\beta}(\partial_{\alpha}\psi)(\partial_{\beta}\psi)
-\frac{1}{4}\tilde{\eta}D_{\;\;\rho\sigma}^{\alpha\beta}\big[(\partial_{\alpha}\psi)(\partial_{\beta}\psi)
\big]&=&0\nonumber\\
\partial_\mu\Big[\Big(\eta^{\mu\nu}+\frac{1}{2}\eta^{\mu\nu}h-h^{\mu\nu}
 \Big)\partial_\nu\psi\Big]+\frac{\tilde{\eta}}{2}
\partial_{\mu}\Big[\big(D^{\mu\nu}_{\;\;\rho\sigma}h^{\rho\sigma}\big)\partial_{\nu}\psi\Big]&=&0.
\end{eqnarray}
It seems that both equations contain more that two time derivatives
acting on fields (the graviton equation contains three time
derivatives acting on the scalar field, while the scalar field
equation contains three time derivatives acting on the graviton).
But in fact they don't. Then we prove it.

We define the following quantities:
\begin{eqnarray}
&&X_{\mu\nu\rho\sigma}=\partial_{(\rho}\eta_{\sigma)(\mu}\partial_{\nu)}
=\frac{1}{4}\big[\eta_{\sigma\mu}\partial_{\rho}\partial_{\nu}
+\eta_{\rho\mu}\partial_{\sigma}\partial_{\nu}
+\eta_{\sigma\nu}\partial_{\rho}\partial_{\mu}
+\eta_{\rho\nu}\partial_{\sigma}\partial_{\mu}\big],\nonumber\\&&
Y_{\mu\nu\rho\sigma}=-\frac{1}{2}\eta_{\mu(\rho}\eta_{\sigma)\nu}
\partial^2=-\frac{1}{4}\big[\eta_{\mu\rho}\eta_{\sigma\nu}
\partial^\tau\partial_\tau+\eta_{\mu\sigma}\eta_{\rho\nu}
\partial^\tau\partial_\tau\big],\nonumber\\&&
Z_{\mu\nu\rho\sigma}=-\frac{1}{2}\eta_{\rho\sigma}\partial_{\mu}\partial_{\nu},\nonumber\\&&
K_{\mu\nu\rho\sigma}=-\frac{1}{2}\eta_{\mu\nu}\partial_\rho\partial_\sigma,\nonumber\\&&
L_{\mu\nu\rho\sigma}=+\frac{1}{2}\eta_{\mu\nu}
\eta_{\rho\sigma}\partial^2=\frac{1}{2}\eta_{\mu\nu}
\eta_{\rho\sigma}\partial^\tau\partial_\tau,
\end{eqnarray}
and firstly calculate the term
$D_{\;\;\rho\sigma}^{\alpha\beta}\big[(\partial_{\alpha}\psi)(\partial_{\beta}\psi)
\big]$.
\begin{eqnarray}
X_{\;\;\rho\sigma}^{\alpha\beta}\big[(\partial_{\alpha}\psi)(\partial_{\beta}\psi)\big]
&=&\frac{1}{2}\psi_{,\rho\sigma}\psi_{,\beta}^{,\beta}
+\frac{1}{4}\psi_{,\rho\alpha}\psi_{,\sigma}^{,\alpha}
+\frac{1}{4}\psi_{,\sigma\alpha}\psi_{,\rho}^{,\alpha}
+\frac{1}{2}\psi_{,\alpha}\psi_{,\rho\sigma}^{,\alpha}
+\frac{1}{4}\psi_{,\sigma}\psi_{,\rho\beta}^{,\beta}
+\frac{1}{4}\psi_{,\rho}\psi_{,\sigma\beta}^{,\beta}\nonumber\\&&
+\frac{1}{2}\psi_{,\rho\sigma}\psi_{,\alpha}^{,\alpha}
+\frac{1}{4}\psi_{,\rho\beta}\psi_{,\sigma}^{,\beta}
+\frac{1}{4}\psi_{,\sigma\beta}\psi_{,\rho}^{,\beta}
+\frac{1}{2}\psi_{,\beta}\psi_{,\rho\sigma}^{,\beta}
+\frac{1}{4}\psi_{,\sigma}\psi_{,\rho\alpha}^{,\alpha}
+\frac{1}{4}\psi_{,\rho}\psi_{,\sigma\alpha}^{,\alpha},\nonumber\\
Y_{\;\;\rho\sigma}^{\alpha\beta}\big[(\partial_{\alpha}\psi)(\partial_{\beta}\psi)\big]&=&
-\frac{1}{2}\big[\psi_{,\rho}^{,\tau}\psi_{,\sigma\tau}
+\psi_{,\sigma}^{,\tau}\psi_{,\rho\tau}+\psi_{,\rho}\psi^{,\tau}_{,\sigma\tau}
+\psi_{,\sigma}\psi^{,\tau}_{,\rho\tau} \big],\nonumber\\
Z_{\;\;\rho\sigma}^{\alpha\beta}\big[(\partial_{\alpha}\psi)(\partial_{\beta}\psi)\big]&=&
-\frac{1}{2}\eta_{\rho\sigma}\big[\psi_{,\alpha}^{,\alpha}\psi_{,\beta}^{,\beta}
+\psi_{,\alpha}\psi_{,\beta}^{,\alpha\beta}
+\psi_{,\alpha}^{,\beta}\psi_{,\beta}^{,\alpha}
+\psi_{,\beta}\psi_{,\alpha}^{,\alpha\beta}\big],\nonumber\\
K_{\;\;\rho\sigma}^{\alpha\beta}\big[(\partial_{\alpha}\psi)(\partial_{\beta}\psi)\big]&=&
-\frac{1}{2}\big[\psi_{,\rho\alpha}\psi_{,\sigma}^{,\alpha}+
\psi_{,\alpha}\psi_{,\rho\sigma}^{,\alpha}+\psi_{,\sigma\alpha}\psi_{,\rho}^{,\alpha}
+\psi_{,\rho\sigma\alpha}\psi^{,\alpha}\big],\nonumber\\
L_{\;\;\rho\sigma}^{\alpha\beta}\big[(\partial_{\alpha}\psi)(\partial_{\beta}\psi)\big]&=&+
\frac{1}{2}\eta_{\rho\sigma}\big[\psi_{,\tau\alpha}\psi^{,\tau\alpha}
+\psi_{,\alpha}\psi_{,\tau}^{,\alpha\tau}
+\psi_{,\alpha}^{,\tau}\psi_{,\tau}^{,\alpha}
+\psi^{,\alpha}\psi^{,\tau}_{,\alpha\tau}\big].
\end{eqnarray}
So we obtain
\begin{eqnarray}
D_{\;\;\rho\sigma}^{\alpha\beta}\big[(\partial_{\alpha}\psi)(\partial_{\beta}\psi)
\big]&=&\frac{1}{4}\eta_{\rho\sigma}\big[\psi_{,\alpha\beta}\psi^{,\alpha\beta}
-\psi_{,\alpha}^{,\alpha}\psi_{,\beta}^{,\beta}\big]-
\frac{1}{4}\big[\psi_{,\rho\alpha}\psi_{,\sigma}^{,\alpha}
+\psi_{,\sigma\alpha}\psi_{,\rho}^{,\alpha}+2\psi_{,\rho\sigma}\psi_{,\alpha}^{,\alpha}\big],
\end{eqnarray}
which do not contain more than two time derivatives acting on the
scalar fields.
 Secondly we calculate the term
$\partial_{\mu}\big(D^{\mu\nu}_{\;\;\rho\sigma}h^{\rho\sigma}\big)$.
\begin{eqnarray}
\partial_{\mu}\big(X^{\mu\nu}_{\;\;\rho\sigma}h^{\rho\sigma}\big)&=&
\frac{1}{4}\Big(h^{\rho\mu,\nu}_{,\mu\rho}+h^{\sigma\mu,\nu}_{,\mu\sigma}
+h^{\rho\nu,\mu}_{,\mu\rho}+h^{\sigma\nu,\mu}_{,\mu\sigma}\Big),\nonumber\\
\partial_{\mu}\big(Y^{\mu\nu}_{\;\;\rho\sigma}h^{\rho\sigma}\big)&=&
-\frac{1}{2}h^{\mu\nu,\tau}_{,\tau\mu},\nonumber\\
\partial_{\mu}\big(Z^{\mu\nu}_{\;\;\rho\sigma}h^{\rho\sigma}\big)&=&
-\frac{1}{2}h^{,\mu\nu}_{,\mu},\nonumber\\
\partial_{\mu}\big(K^{\mu\nu}_{\;\;\rho\sigma}h^{\rho\sigma}\big)&=&
-\frac{1}{2}h^{\rho\sigma,\nu}_{,\rho\sigma},\nonumber\\
\partial_{\mu}\big(L^{\mu\nu}_{\;\;\rho\sigma}h^{\rho\sigma}\big)&=&
\frac{1}{2}h^{,\tau\nu}_{,\tau}.
\end{eqnarray}
Then we obtain
\begin{eqnarray}
\partial_{\mu}\big(D^{\mu\nu}_{\;\;\rho\sigma}h^{\rho\sigma}\big)=0,
\end{eqnarray}
which don't contain three time derivatives acting on the graviton.
Lastly we find the motion and graviton equations are
\begin{eqnarray}
 -\frac{1}{16\pi
G}D_{\mu\nu\rho\sigma}h^{\rho\sigma}+\frac{1}{2}(\partial_{\mu}\psi)(\partial_{\nu}\psi)
-\frac{1}{4}\eta_{\mu\nu}\eta^{\alpha\beta}(\partial_{\alpha}\psi)(\partial_{\beta}\psi)&&\nonumber\\
-\frac{\tilde{\eta}}{16}\Big\{\eta_{\rho\sigma}\big[\psi_{,\alpha\beta}\psi^{,\alpha\beta}
-\psi_{,\alpha}^{,\alpha}\psi_{,\beta}^{,\beta}\big]-
\big[\psi_{,\rho\alpha}\psi_{,\sigma}^{,\alpha}
+\psi_{,\sigma\alpha}\psi_{,\rho}^{,\alpha}+2\psi_{,\rho\sigma}\psi_{,\alpha}^{,\alpha}\big]\Big\}&=&0,
\nonumber\\
\partial_\mu\Big[\Big(\eta^{\mu\nu}+\frac{1}{2}\eta^{\mu\nu}h-h^{\mu\nu}
 \Big)\partial_\nu\psi\Big]+\frac{\tilde{\eta}}{2}
\big(D^{\mu\nu}_{\;\;\rho\sigma}h^{\rho\sigma}\big)\partial_{\mu}\partial_{\nu}\psi&=&0.
\end{eqnarray}
 Note that neither the graviton nor the scalar equation of motion
contain three time derivatives, suggesting that these type of
theories may have, after all, energy bounded from below, and hence
represent physically acceptable theories.


\vspace*{0.2cm}
\begin{thebibliography}{99}
\baselineskip=0.6 cm

\bibitem{chensong}S. Chen, B. Wang and R. Su, {\it Hawking radiation in a $d$-dimensional static spherically-symmetric black Hole surrounded by quintessence } Phys. Rev. D 77, 124011 (2008).

\bibitem{inf1} A. H. Guth, {\it Inflationary universe: A possible solution to the horizon and flatness problems} Phys. Rev. D 23, 347 (1981).

\bibitem{1a} B. Ratra and J. Peebles, {\it Cosmological Consequences Of A Rolling Homogeneous Scalar Field } Phys. Rev. D 37,  3406 (1988);
C. Wetterich, {\it Cosmology And The Fate Of Dilatation Symmetry} Nucl. Phys. B 302, 668 (1988); R. R. Caldwell,
R. Dave and P. J. Steinhardt, {\it Cosmological Imprint of an Energy Component with General Equation of State } Phys. Rev. Lett. 80, 1582 (1998
);  M. Doran and J. Jaeckel, {\it Loop Corrections to Scalar Quintessence Potentials } Phys. Rev. D 66, 043519 (2002).

\bibitem{2a} C. A.  Picon, T. Damour and V. Mukhanov, {\it k-Inflation } Phys. Lett.
B 458, 209 (1999); T. Chiba, T. Okabe and M. Yamaguchi, {\it Kinetically Driven Quintessence } Phys.
Rev. D 62 023511 (2000).

\bibitem{3a} R. R. Caldwell, {\it A Phantom Menace? Cosmological consequences of a dark energy component with super-negative equation of state } Phys. Lett. B 545, 23 (2002);
B. McInnes, {\it The dS/CFT Correspondence and the Big Smash } J. High Energy Phys. 08, 029 (2002); S. Nojiri and
S. D. Odintsov, {\it Quantum deSitter cosmology and phantom matter } Phys. Lett. B 562, 147 (2003);  L. P. Chimento
and R. Lazkoz, {\it Constructing Phantom Cosmologies from Standard Scalar Field Universes } Phys. Rev. Lett. 91, 211301 (2003); B.
Boisseau, G. Esposito-Farese, D. Polarski, Alexei A.
Starobinsky, {\it Reconstruction of a scalar-tensor theory of gravity in an accelerating universe } Phys. Rev. Lett. 85, 2236 (2000); R. Gannouji, D.
Polarski, A. Ranquet, A. A. Starobinsky, {\it Scalar¨Ctensor models of normal and phantom dark energy } JCAP 0609, 016 (2006).

\bibitem{Hig} P. W. Higgs, Phys. Lett. B {\bf12}, 132 (1964).

\bibitem{bd} C. Brans and R. H. Dicke, {\it Mach's Principle and a Relativistic Theory of Gravitation} Phys. Rev. 124, 925 (1961).

\bibitem{KK} C. Csaki, M. Graesser, L. Randall, J. Terning, {\it Cosmology of Brane Models with Radion Stabilization } Phys. Rev. D 62,
045015 (2000).

\bibitem{dila} G. W. Gibbons, K. Maeda, {\it Black Holes And Membranes In Higher Dimensional Theories With Dilaton Fields} Nucl. Phys. B 298, 741
(1988).

\bibitem{kkc} Q. Shafi and C. Wetterich, {\it Inflation With Higher Dimensional Gravity} Phys. Lett. B 152, 51
(1985).

\bibitem{kkc1} Q. Shafi and C.Wetterich, {\it Inflation From Higher Dimensions
} Nucl. Phys. B 289, 787
(1987).

\bibitem{kkc2} A. Linde, \textit{Particle Physics and Inflationary Cosmology}, (1990) Harwood Publ.
London.

\bibitem{AL1} L. Amendola, {\it Cosmology with Nonminimal Derivative Couplings } Phys. Lett. B 301, 175 (1993).


\bibitem{AL3} S. Capozziello, G. Lambiase, {\it Nonminimal Derivative Coupling and the Recovering of Cosmological Constant } Gen. Rel. Grav. 31,  1005 (1999).

\bibitem{AL4} S. F. Daniel and R. Caldwell, {\it Consequences of a Cosmic Scalar with Kinetic Coupling to Curvature } Class. Quant. Grav. 24, 5573 (2007).

\bibitem{AL5} S. V. Sushkov, {\it Exact cosmological solutions with nonminimal derivative coupling } Phys. Rev. D 80, 103505 (2009).

\bibitem{g1} C. J. Gao, {\it When scalar field is kinetically coupled to the Einstein tensor } JCAP 06 (2010) 023.

\bibitem{g2} L.N. Granda, {\it Non-minimal Kinetic coupling to gravity and accelerated expansion } arXiv: 0911.3702.

\bibitem{g3} E. N. Saridakis and S. V. Sushkov, {\it Quintessence and phantom cosmology with non-minimal derivative coupling } Phys. Rev. D 81, 083510, (2010).
\bibitem{songbai} S. B. Chen and J. L. Jing, {\it Greybody factor for a scalar field coupling to Einstein's tensor } Phys. Lett. B 691, 254 (2010).
 \bibitem{shir}T. Shiromizu and U. Gen, {\it A Probe Particle in Kerr-Newman-deSitter Cosmos}, Class. Quant. Grav. 17, 1361 (2000).
\bibitem{Kanti} P. Kanti, R. A. Konoplya, A. Zhidenko, {\it Quasi-Normal Modes of Brane-Localised Standard Model Fields II: Kerr Black Holes } Phys. Rev. D 74, 064008 (2006);
P. Kanti, R. A. Konoplya, {\it Quasi-Normal Modes of Brane-Localised Standard Model Fields } Phys. Rev. D 73, 044002 (2006); D. K.
Park, {\it Asymptotic Quasinormal Frequencies of Brane-Localized Black Hole } Phys. Lett. B 633, 613 (2006).

\bibitem{Kan1} P. Kanti, {\it Reading the Number of Extra Dimensions in the Spectrum of Hawking Radiation } hep-ph/0310162.

\bibitem{Kan2} C. M. Harris and P. Kanti, {\it Hawking Radiation from a (4+n)-dimensional Black Hole: Exact Results for the Schwarzschild Phase } JHEP 0310 014 (2003) ; P. Kanti, {\it Black Holes in Theories with Large Extra Dimensions: a Review } Int. J. Mod. Phys. A 19 4899 (2004);
P. Argyres, S. Dimopoulos and J. March-Russell, {\it Black Holes and Sub-millimeter Dimensions } Phys. Lett.
B 441, 96 (1998); T. Banks and W. Fischler, {\it A Model for High Energy Scattering in Quantum Gravity } hep-th/9906038; R.
Emparan, G. T. Horowitz and R. C. Myers, {\it Black Holes Radiate Mainly on the Brane } Phys. Rev. Lett. 85,
499 (2000).

\bibitem{Kan3} E. Jung and D. K. Park, {\it Absorption and Emission Spectra of an higher-dimensional Reissner-Nordstr\"{o}m black hole
} Nucl. Phys. B 717, 272 (2005); N. Sanchez, {\it Absorption And Emission Spectra Of A Schwarzschild Black Hole} Phys. Rev. D 18, 1030 (1978);
E. Jung and D. K. Park, {\it Effect of Scalar Mass in the Absorption and Emission Spectra of Schwarzschild Black Hole } Class. Quant. Grav. 21, 3717 (2004); E.
Jung, S. H. Kim and D. K. Park, {\it Low-Energy Absorption Cross Section for massive scalar and Dirac fermion by $(4+n)$-dimensional Schwarzschild Black Hole } JHEP 0409 (2004) 005.

\bibitem{Kan4} A. S. Majumdar, N. Mukherjee, {\it Braneworld black holes in cosmology and astrophysics } Int. J. Mod. Phys. D 14 1095 (2005) and reference therein;
G. Kofinas, E. Papantonopoulos and V. Zamarias, {\it Black Hole Solutions in Braneworlds with Induced Gravity } Phys. Rev. D
66, 104028 (2002); G. Kofinas, E. Papantonopoulos and V.
Zamarias, {\it Black Holes on the Brane with Induced Gravity } Astrophys. Space Sci. 283, 685 (2003); A. N. Aliev,
A. E. Gumrukcuoglu, {\it Charged Rotating Black Holes on a 3-Brane } Phys. Rev. D 71, 104027 (2005); S. Kar, S.
Majumdar, {\it Black Hole Geometries in Noncommutative String Theory } Int. J. Mod. Phys. A 21, 6087 (2006) ; S. Kar, S.
Majumdar, {\it Noncommutative $D_3$-brane, Black Holes and Attractor Mechanism } Phys. Rev. D 74, 066003 (2006); S. Kar, {\it Tunneling between de Sitter and anti de Sitter black holes in a noncommutative D3-brane formalism
} Phys. Rev. D
74, 126002 (2006).

\bibitem{Kan5} E. Jung, S. H. Kim and D. K. Park, {\it Condition for Superradiance in Higher-dimensional Rotating Black Holes } Phys. Lett. B 615, 273 (2005);
E. Jung, S. H. Kim and D. K. Park, {\it Condition for the Superradiance Modes in Higher-Dimensional Rotating Black Holes with Multiple Angular Momentum Parameters } Phys. Lett. B 619, 347
(2005); D. Ida, K. Oda and S. C. Park, {\it Rotating black holes at future colliders: Greybody factors for brane fields } Phys. Rev. D 67, 064025
(2003); G. Duffy, C. Harris, P. Kanti and E. Winstanley, {\it Dynamics of Quark-Gluon-Plasma Instabilities in Discretized Hard-Loop Approximation } JHEP
0509, 041 (2005); M. Casals, P. Kanti and E. Winstanley, {\it Brane Decay of a (4+n)-Dimensional Rotating Black Hole. II: spin-1 particles } JHEP
0602, 051 (2006); E. Jung and D. K. Park, {\it Bulk versus Brane in the Absorption and Emission : 5D Rotating Black Hole Case } Nucl. Phys. B 731,
171 (2005); A. S. Cornell, W. Naylor and M. Sasaki, {\it Graviton emission from a higher-dimensional black hole } JHEP 0602,
012 (2006); V. P. Frolov, D. Stojkovic, {\it Black Hole as a Point Radiator and Recoil Effect on the Brane World } Phys. Rev. Lett. 89,
151302 (2002); Valeri P. Frolov, Dejan Stojkovic, {\it Black Hole Radiation in the Brane World and Recoil Effect } Phys. Rev. D
66, 084002 (2002); D. Stojkovic, {\it Distinguishing between the small ADD and RS black holes in accelerators } Phys. Rev. Lett. 94,
011603 (2005).

\bibitem{Kan6} Eylee Jung and D. K. Park, {\it Validity of Emparan-Horowitz-Myers argument in Hawking radiation into massless spin-2 fields
} Mod. Phys. Lett. A, 22, 1635 (2007); V. Cardoso, M. Cavaglia, L.
Gualtieri, {\it Black hole particle emission in higher-dimensional
spacetimes } Phys. Rev. Lett. 96, 071301 (2006); V. Cardoso, M.
Cavaglia, L. Gualtieri, {\it Hawking emission of gravitons in higher
dimensions: non-rotating black holes } JHEP 0602, 021 (2006).

\bibitem{Haw3} S. Creek, O. Efthimiou, P. Kanti and K. Tamvakis, {\it Greybody Factors for Brane Scalar Fields in a Rotating Black-Hole Background }  Phys. Rev. D 75, 084043
(2007).

\bibitem{Haw4}S. B. Chen, B. Wang and R.-K. Su, {\it Hawking radiation in a Rotating Kaluza-Klein Black Hole with Squashed Horizons } Phys. Rev. D 77, 024039
(2008); {\it Greybody Factors for Rotating Black Holes on Codimension-2 Branes } JHEP.0803, 019 (2008);  S. B. Chen, B. Wang and J. L.
Jing, {\it Scalar emission in a rotating G?del black hole } Phys. Rev. D 78, 064030 (2008).


\bibitem{Haw5}C. M. Harris and P. Kanti, {\it Hawking Radiation from a (4+n)-Dimensional Rotating Black Hole on the Brane } Phys. Lett. B 633, 106 (2006); D. Ida, K.
Oda and S. C. Park, {\it Rotating black holes at future colliders II: Anisotropic scalar field emission } Phys. Rev. D 71, 124039 (2005);

\bibitem{ding} C. K. Ding and J. L. Jing, {\it Deformation of contour and Hawking temperature } Class. Quant. Grav. 27, 035004 (2010).
\bibitem{mb} M. Abramowitz and I. Stegun, \textit{Handbook of Mathematical
Functions} (Academic, New York, 1996).


\end{thebibliography}
\end{document}